\documentclass[preprint,pre]{revtex4-1}

\usepackage[utf8]{inputenc}
\usepackage[english]{babel}
\usepackage{amsmath}
\usepackage{amssymb}
\usepackage{latexsym}
\usepackage{graphicx}
\usepackage{color}
\usepackage[dvipsnames]{xcolor}

\renewcommand{\v}[1]{\ensuremath{\mathbf{#1}}}
\newcommand{\pd}[2]{\frac{\partial #1}{\partial #2}}
\providecommand{\vr}{\v{r}}
\renewcommand{\vr}{\v{r}}

\newcommand{\vF}{\v{F}}
\newcommand{\vk}{\v{k}}
\newcommand{\vx}{\v{x}}
\newcommand{\vs}{\v{s}}
\newcommand{\vp}{\v{p}}
\newcommand{\rd}{{\rm d}}
\newcommand{\ii}{\mathrm{i}}
\newcommand{\ee}{\mathrm{e}}
\newcommand{\vnabla}{\boldsymbol{\nabla}}
\newcommand{\vcdot}{\boldsymbol{\cdot}}

\begin{document}

\date{August 7, 2017}
\title{Efficient solution of the Wigner-Liouville equation using a spectral decomposition of the force field}
\author{Maarten L. \surname{Van de Put}}
\email[Electronic mail: ]{vandeput.maarten@gmail.com}
\affiliation{Department~of~Physics, Universiteit~Antwerpen, B-2020~Antwerpen, Belgium}
\affiliation{imec, B-3001~Leuven, Belgium}
\author{Bart Sor\'ee}
\altaffiliation{Department of Electrical Engineering (ESAT), KU~Leuven, B-3001~Leuven, Belgium}
\affiliation{Department~of~Physics, Universiteit~Antwerpen, B-2020~Antwerpen, Belgium}
\affiliation{imec, B-3001~Leuven, Belgium}
\author{Wim Magnus}
\affiliation{Department~of~Physics, Universiteit~Antwerpen, B-2020~Antwerpen, Belgium}
\affiliation{imec, B-3001~Leuven, Belgium}

\begin{abstract}
The Wigner-Liouville equation is reformulated using a spectral decomposition of the classical force field instead of the potential energy.
The latter is shown to simplify the Wigner-Liouville kernel both conceptually and numerically as the spectral force Wigner-Liouville equation avoids the numerical evaluation of the highly oscillatory Wigner kernel which is nonlocal in both position and momentum.
The quantum mechanical evolution is instead governed by a term local in space and non-local in momentum, where the non-locality in momentum has only a limited range.
An interpretation of the time evolution in terms of two processes is presented;  a classical evolution under the influence of the averaged driving field, and a probability-preserving quantum-mechanical generation and annihilation term.
Using the inherent stability and reduced complexity, a direct deterministic numerical implementation using Chebyshev and Fourier pseudo-spectral methods is detailed.
For the purpose of illustration, we present results for the time-evolution of a one-dimensional resonant tunneling diode driven out of equilibrium.
\end{abstract}

\maketitle

\section{Introduction}

The investigation of electronic transport in electronic devices takes place at a crossroads of computational physics, quantum mechanics and statistical mechanics.
With the advent of nanoscaled components, quantum mechanical models replace the older semi-classical models.
Nowadays, there are several competing frameworks for calculating the quantum transport models, ranging from simple single particle ballistic wavefunction based methods \cite{Lent:1990fg} over the non-equilibrium Green's functions (NEGF) methods \cite{Datta:1997tk} to the Pauli master equation \cite{Fischetti:1999df,Fischetti:1998gt} and the Wigner formalism \cite{Wigner:1932cz,Jacoboni:2004gf}.

In the Wigner formalism, one describes a generally mixed quantum state expressed in terms of a Wigner function $f(\vr,\vp,t)$ living in phase space.
Being a quasi-distribution function, the Wigner function acts like a distribution function for the purpose of calculating real observables, but it cannot be used as a real probability function because it can, and does, attain negative values in the presence of quantum mechanical effects \cite{Kenfack:2004cv}.
Like the semi-classical Boltzmann transport equation, the Wigner formalism is not limited to calculating the steady state, but can be used to calculate the full time evolution of a state $f(\vr,\vp,t)$, under the influence of a generally time-dependent potential profile $V(\vr,t)$, starting from a given initial state, such as the equilibrium state.
Furthermore, technology has progressed to the point that very fast pulsing in the range of picoseconds is available, and already used in the manipulation of charge based quantum well qubits in quantum computing \cite{Aassime:2001ig,Hollenberg:2004ge}.
The Wigner function formalism is naturally equipped to handle the time-evolution, including mixing of states and decoherence, in these cases.

The quantum mechanical analogue to the classical Boltzmann transport equation, describing the time evolution of the Wigner function is the Wigner-Liouville equation,
\begin{equation}
  \pd{}{t} f(\vr,\vp,t)
  + \frac{\vp}{m} \vcdot \vnabla_\vr f(\vr,\vp,t)
  - \frac{1}{\hbar} \int\rd^3p'\, W(\vr,\vp',t) f(\vr,\vp-\vp',t) = 0\,,
  \label{e:wl}
\end{equation}
where the Wigner kernel is given by
\begin{equation}
  W(\vr,\vp,t) = {-}\frac{\ii}{(2\pi\hbar)^3} \int\rd^3x \,\ee^{{-}\ii\vp\vcdot\vx/\hbar}
                \left[ V\!\left(\vr+\frac{\vx}{2},t\right) - V\!\left(\vr-\frac{\vx}{2},t\right) \right]\,,
  \label{e:wl-kernel}
\end{equation}
where we have omitted the Boltzmann collision integral for the time being.
Unlike its classical counterpart, the influence of the potential on the Wigner function through the Wigner kernel is non-local in both position and momentum.
This non-locality, combined with the high dimensionality of phase space - typically three times the dimension of the configuration space, makes the Wigner kernel difficult to handle directly.

In literature, we find a range of proposed techniques to tackle this problem.
In line with the solution of the Boltzmann equation, there have been significant efforts in solving the Wigner equation using Monte-Carlo, or particle based techniques \cite{Sellier:2015ef,Jonasson:2015ew,Jacoboni:2004gf,Ellinghaus:2014ef,Weinbub:2015fq}.
Even though all Monte-Carlo approaches rely on the stochastic solution of the Wigner function, the differences between various Monte-Carlo approaches are significant, from relying on signed particles \cite{Sellier:2015ef} and Fourier transforms  \cite{Jonasson:2015ew} to using spatial decomposition approaches  \cite{Ellinghaus:2014ef,Weinbub:2015fq}.
In contrast to the former, the deterministic approaches feature a more direct treatment, which we have adopted in this paper.
While having been studied for a long time \cite{Frensley:1990by}, direct discretization using well-designed finite difference approaches has recently been improved by using the weighted essentially non-oscillating scheme (WENO) to prevent erroneous oscillations \cite{Dorda:2015gu}.
In other work, it has also been shown that the Wigner function can be solved in a robust way using adaptive pseudo-spectral methods in the form of mass-conserving spectral element methods \cite{shao:2011xx,xiong:2016xx}. In these methods, the carefully crafted spectral decomposition of the Wigner function enables the oscillatory components introduced by the Wigner kernel to be solved exactly.
Finally, the oscillatory quantum effects can also be mitigated by decomposing the potential in a classical and quantum part \cite{Sellier:2015gz,Sels:2012kc}.

In this paper, we propose a deterministic method to solve the Wigner equation by exploiting the analogy with the classical Boltzmann transport equation without, however, invoking the classical limit.
In contrast to the highly optimized numerical techniques that have been outlined above \cite{Sellier:2015ef,Jonasson:2015ew,Jacoboni:2004gf,Ellinghaus:2014ef,Weinbub:2015fq,Dorda:2015gu,shao:2011xx,xiong:2016xx}, our main focus is on the development of a different, but equivalent form of the Wigner function that allows for a natural, direct implementation.
To this end, we recast the Wigner equation into a form that employs the force field rather than the potential energy from which one would conventionally proceed in quantum mechanics.
In section~\ref{force}, we detail the reformulation of the Wigner equations using spectral components of the force field.
In section~\ref{1d} we consider the one-dimensional case and its interpretation based on signed generation terms.
In section~\ref{implementation}, we describe a one-dimensional numerical implementation of the rewritten Wigner equation using Chebyshev and Fourier pseudo-spectral methods.
In section~\ref{rtd}, we illustrate our method by tracing the time evolution of electrons propagating through a double-barrier resonant tunneling diode (RTD) driven out of equilibrium.
Finally, we conclude in section~\ref{conclusions}.

\section{Using the classical force}
\label{force}

Through its appearance in the Wigner kernel, the potential energy $V(\vr)$ describes the non-local change in momentum of the quasi-distribution function.
This is contrasted by the classical counterpart of the Wigner-Liouville equation, the Boltzmann transport equation, where the classical force field $\vF(\vr,t) = -\vnabla V(\vr,t)$ provides a direct, local measure of the change in the momentum coordinate over time.
In this section, we recast the quantum mechanical Wigner-Liouville equation in terms of the classical force.

As $V(\vr)$ enters the Wigner kernel merely through the potential difference between the points $\vr-\vx/2$ and $\vr+\vx/2$, we express the latter as a line integral of the force along a path $\Gamma$ from $\vr-\vx/2$ to $\vr+\vx/2$,
\begin{equation}
  V\left(\vr+\frac{\vx}{2}\right) - V\left(\vr-\frac{\vx}{2}\right)
  = -\int\limits_\Gamma \rd\vs \vcdot \vF(\vs) \,.
\label{e:VtoFlineint}
\end{equation}
Note that, because $\vF(\vs)$ is a conservative field, $\Gamma$ can be any continuous curve connecting $\vr-\vx/2$ to $\vr+\vx/2$.

Next, we accommodate for the wave-like nature of quantum mechanics by decomposing the force in its plane wave components using the Fourier transform,
\begin{equation}
  \vF(\vr) = \int\rd^3k\, \ee^{\ii\vk\vcdot\vr}\, \tilde{\vF}(\vk) \,.
\label{e:FtoFk}
\end{equation}
Combining (\ref{e:VtoFlineint}) and (\ref{e:FtoFk}), we obtain a line integral over plane waves,
\begin{equation}
  \int\rd^3k\, \tilde{\vF}(\vk)\vcdot \int\limits_\Gamma \rd\vs\, \ee^{\ii\vk\vcdot\vs} \,,
  \label{e:Fklineint}
\end{equation}
which can be evaluated by taking $\Gamma$ to be the straight line $\vs(\alpha) = \vr + \alpha\vx$  with $\alpha$ running from $-1/2$ to $1/2$, yielding
\begin{equation}
  \int\rd^3k\, \tilde{\vF}(\vk) \ee^{\ii\vk\vcdot\vr}\vcdot \vx
  \int\limits_{-1/2}^{1/2} \rd\alpha\, \ee^{\ii\alpha\vk\vcdot\vx} \,.
  \label{e:Fkstraightlineint}
\end{equation}

Using $\vnabla_\vp \ee^{\ii\vp\vcdot\vx/\hbar} = \ii\vx/\hbar \ee^{\ii\vp\vcdot\vx/\hbar}$ and integrating over the relative coordinate $\vx$, we reduce the Wigner kernel (\ref{e:wl-kernel}) to
\begin{equation}
  W(\vr,\vp) = \int\rd^3k\, \vF_\vk(\vr) \vcdot \vnabla_\vp
                 \int\limits_{-1/2}^{1/2} \rd\alpha\, \delta(\alpha\hbar\vk - \vp)\,,
  \label{e:wl-kernel-reduced}
\end{equation}
where $\delta(\vp)=\delta(p_x)\delta(p_y)\delta(p_z)$ is the three-dimensional Dirac delta function and $\vF_\vk(\vr) = \tilde{\vF}(\vk) \ee^{\ii\vk\vcdot\vr}$ are the spectral components of the force.

Inserting this in the Wigner-Liouville equation, we may eliminate the momentum integral to arrive at
\begin{equation}
  \pd{}{t} f(\vr,\vp,t)
  + \frac{\vp}{m} \vcdot \vnabla_\vr f(\vr,\vp,t)
  + \int\rd^3k\, \vF_\vk(\vr) \vcdot \vnabla_\vp \int\limits_{-1/2}^{1/2} \rd\alpha\, f(\vr,\vp-\alpha\hbar\vk,t) = 0\,.
  \label{e:wl-spectral}
\end{equation}

Being fully equivalent with the standard Wigner-Liouville equation (\ref{e:wl}), equation~(\ref{e:wl-spectral}) is  built on the spectral components of the force field $\vF_\vk(\vr,t)$ rather than the potential energy, thereby revealing some interesting features.
Firstly, (\ref{e:wl-spectral}) is very similar in form to the classical Boltzmann equation which facilitates the physical interpretation of the Wigner-Liouville equation.
It readily reduces to the classical limit if the spectral components of the force are only significant for wave vectors $\vk$ that are smaller than the characteristic size at which the Wigner function changes in the momentum coordinate $\vp$.
Secondly, the Fourier transformation has absorbed all non-local interactions in the position coordinate.
Finally, even though  the interaction in momentum space is non-local, as is to be expected when quantum mechanics rules, its range remains finite, being limited by $\vp\pm\hbar\vk/2$.
As such, for realistic potential profiles, where effects like screening limit the amplitude of Fourier components of the potential at large wave vectors, the smoothness of the corresponding force field introduces a natural limit on the wave vector components of the force, which in turn limits the non-locality in momentum space. Note that this natural limit is problem dependent, and that one can construct toy models with non-continuous potential profiles where such a limit does not exist in principle.

While these properties are valuable on their own, we believe that the biggest benefit of using the spectral decomposition of the force field is that it allows for a straightforward numerical treatment.
In particular, the semi-local interaction in (\ref{e:wl-spectral}) can be directly computed a lot easier than its non-local, equivalent counterpart (\ref{e:wl}), at the marginal cost of a fast Fourier transform that generates the plane wave components of the force field.
Furthermore, all extraneous highly oscillatory terms of the original formulation are removed in (\ref{e:wl-spectral}), except for the plane wave term hidden in the spectral components in $\vF_\vk(\vr)$. Therefore, the fastest oscillations are determined by the force field itself. For a range of realistic, smooth force fields, we have argued that we can apply a natural short wavelength cutoff which suppresses the high wavevector oscillations.
As such, the accuracy of the numerical scheme needed to solve (\ref{e:wl-spectral}) is determined solely by the physical constraints of the system, as would be the case for the classical Boltzmann equation.

\section{The one dimensional case}
\label{1d}

In six-dimensional phase space, the integral over $\alpha$ and the gradient with respect to momentum $\vp$ do not readily reduce to an even simpler form due to the directional component of the wave vector $\vk$.
However, in the case of one-dimensional transport, the gradient reduces to an ordinary partial derivative with respect to momentum, yielding, upon a change of coordinates
\footnote{Strictly speaking, the change of variables $\alpha\hbar k\to p'$ requires a Cauchy principal value integral for $k$ due to a pole at $k=0$. However, as we will make clear later, this is of no consequence to its practical use.}
\begin{equation}
  \pd{}{t} f(x,p,t)
  + \frac{p}{m} \partial_x f(x,p,t)
  + \int\rd k\, \frac{F_k(x)}{\hbar k} \int\limits_{-\hbar k/2}^{\hbar k/2} \rd p'\, \partial_p f(x,p-p',t) = 0\,.
  \label{e:wl-spectral-1d-a}
\end{equation}
Using $\vnabla_\vp \ee^{\ii\vp\vcdot\vx/\hbar} = \ii\vx/\hbar \ee^{\ii\vp\vcdot\vx/\hbar}$ and integrating over the relative coordinate $\vx$, we reduce the Wigner kernel (\ref{e:wl-kernel}) to
\begin{equation}
  \pd{}{t} f(x,p,t)
  + \frac{p}{m} \partial_x f(x,p,t)
  + \int\rd k\, \frac{F_k(x)}{\hbar k} 
  \left[ f\!\left(x,p+\frac{\hbar k}{2},t\right) - f\!\left(x,p-\frac{\hbar k}{2},t\right) \right] = 0\,.
  \label{e:wl-spectral-1d}
\end{equation}
Apparently, only the integral involving the  spectral components of the force field remains.
Its numerical evaluation can safely be restricted to a finite range of $k$-values because both the $k^{-1}$ term and the finite spectrum of a smoothly varying force field provide a natural cut-off.

We interpret equation (\ref{e:wl-spectral-1d}) in terms of a classical ``free'' evolution as well as local, quantum mechanical generation and annihilation events.
The classical evolution is related to the first two terms of (\ref{e:wl-spectral-1d}).
The corresponding trajectory through phase space is a straight line of constant momentum if no other terms were present.
Close inspection of the last term of the left hand side in (\ref{e:wl-spectral-1d}) however reveals an additional classical force term arising for $k=0$ from Newton's difference quotient,
\begin{equation}
	\lim_{k\to0} \frac{F_k(x,t)}{\hbar k}\,
		\left[ f\left(x,p-\frac{\hbar k}{2},t\right) - f\left(x,p+\frac{\hbar k}{2},t\right) \right] 
  = -F_{k=0}(t) \pd{}{p} f(x,p,t).
\end{equation}
Being the $k=0$ Fourier component, $F_{k=0}(t)$ is uniform along the active area, thereby representing the spatially averaged force field.
Consequently, the classical part of (\ref{e:wl-spectral-1d}) describes the evolution of uniformly accelerated particles subjected to a force $F_{k=0}(t)$, while tracing a parabolic trajectory in phase space.

\begin{figure}[!ht]
	\includegraphics{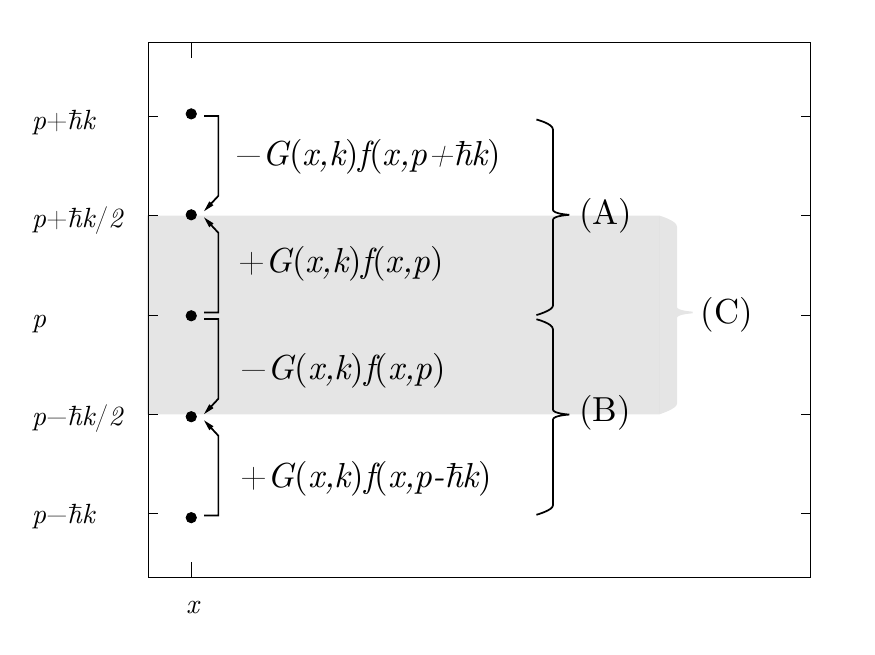}
	\caption{Combination of two generation processes (A) and (B) that don't conserve the particle number into a particle conserving process (C). See text for discussion.}
	\label{f:generation}
\end{figure}

The remaining parts of (\ref{e:wl-spectral-1d}) can be interpreted  as a continuum of local quantum generation and annihilation processes.
The generic transition rate determining the generation and annihilation processes, are given by
\begin{equation}
  G(x,k,t) = \frac{F_k(x,t)}{\hbar k}\,.
\end{equation}
In order to better understand the latter, we study a generation event occurring at a single, fixed wave vector $k$.
According to (\ref{e:wl-spectral-1d}), a positive (negative) value for $F_k(x)$ contributes positively (negatively) from a lower momentum $p-\hbar k/2$, whereas the higher momentum $p+\hbar k/2$ contributes negatively (positively).
While the generation rates are independent of momentum $p$, the actual contribution from lower and higher momenta is generally different as it is  weighed by the value of the Wigner function at that point, as illustrated in figure~\ref{f:generation} for two separate generation and annihilation processes (A) and (B) at respective  momenta $p+\hbar k/2$ and $p-\hbar k/2$.
A single generation and annihilation process with a fixed value of $x, p$ and $k$ does not conserve probability density.
However, by carefully choosing two different points in phase space with a momentum spacing of $\hbar k$, as we have done for the processes (A) and (B), a symmetric generation and annihilation process emerges, denoted by (C) figure~\ref{f:generation}.
This symmetric process launches transitions to higher and lower momenta from a central momentum, conserving probability density by construction.

This symmetric process of negative and positive contributions (generation and annihilation) has been described previously in the context of a Wigner Monte Carlo technique in Refs.~ \cite{Nedjalkov:2008ed,Sellier:2013ga}, where the discrete nature of the Monte Carlo technique has led to a pictorial description in terms of positively and negatively signed particles.
For a more rigorous explanation of these signed particle methods, we refer the reader to refs. \cite{muscato:2016xx,nedjalkov:2013xx}.

In this light, the Wigner-Liouville approach based on the spectral force decomposition could be considered a deterministic counterpart to the stochastic Monte-Carlo approach.
Though being conceptually rich, the latter has proven to be fast, especially when dealing with space and momentum confined distributions such as wave packets.
It is remarkable that, while the approach detailed in this paper is very dissimilar from the MC particle method, a common interpretation of the generation process can be found.

\section{Numerical Implementation}
\label{implementation}

The increased stability arising from the spectral force decomposition of the Wigner-Liouville equation allows for a direct deterministic numerical discretization, as opposed to the stochastic Monte-Carlo approach commonly used to implement the Wigner and Boltzmann equations.

\subsection{Time evolution and discretization}

\begin{figure}[!ht]
  \includegraphics[width=.4\textwidth]{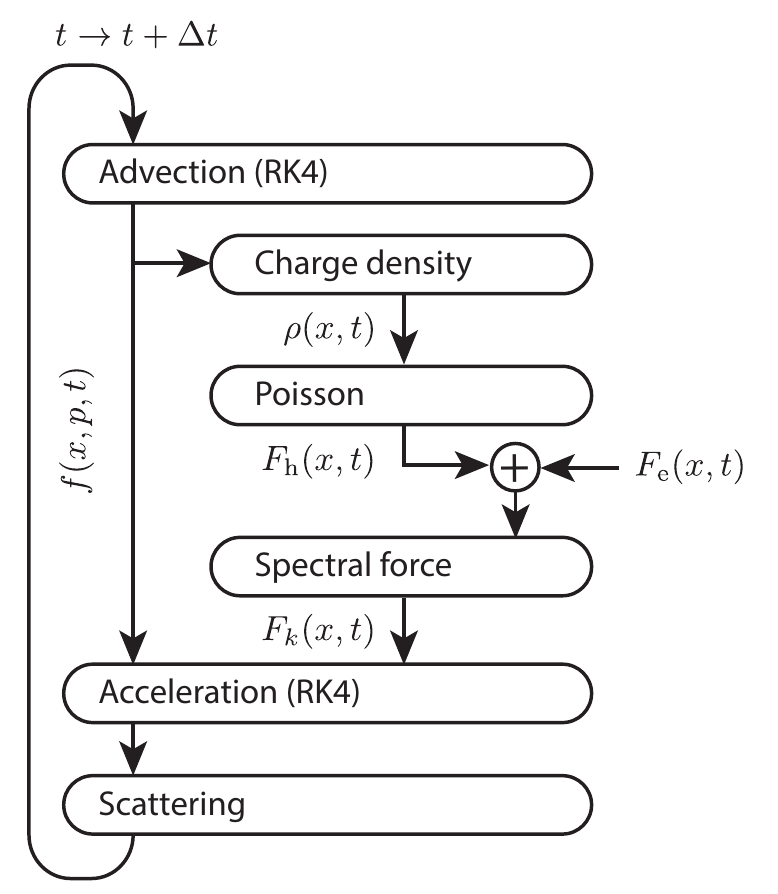}
  \caption{Flow diagram of full time-evolution code}
  \label{f:flow}
\end{figure}

We have implemented the one-dimensional spectral force based Wigner-Liouville equation (\ref{e:wl-spectral-1d}) in a more general time-evolution code. A simplified flow chart of the time marching in this code is outlined in Fig.~\ref{f:flow}.
We use a split-step scheme consisting of an advection, an acceleration and a scattering part.
The solution to Poisson's equation enters the code as a self-consistent update of the force accounting for the charge density changes.
Together, the advection and acceleration parts compose the solution to the Wigner-Liouville equation.
The scattering part is approximated by adding a Boltzmann-like scattering term to the RHS \cite{Sels:2013cv}.
For the purpose of this paper, we merely report on the numerical implementation of the advection and acceleration parts.

The numerical implementation closely follows the interpretation of the generation events explained in section~\ref{1d}.
The advection part involves the change in position, i.e. classical free flow,
\begin{equation}
  \left. \pd{}{t} f(x,p,t) \right|_\textrm{adv} = - \frac{p}{m} \pd{}{x} f(x,p,t)\,,
\end{equation}
while the acceleration part is a combination of the uniform acceleration due to the driving field $F_{k=0}(t)$ and the quantum generation terms,
\begin{equation}
  \left. \pd{}{t} f(x,p,t) \right|_\textrm{accel}
  = - F_{k=0}(t) \pd{}{p} f(x,p,t) + \sum_k F_k(x,t) 
  \left[ f\left(x,p-\frac{\hbar k}{2},t\right) - f\left(x,p+\frac{\hbar k}{2},t\right) \right] \,.
\end{equation}

Both parts calculate the time-evolution over a time step $\Delta t$ invoking the Runge-Kutta method of order 4 (RK4) in turn.
The discretization grids for $p$ and $k$ are chosen such that the quantum generation can be computed without interpolation, the regular spacing being chosen according to $2 \Delta p = \hbar \Delta k$.
Although the discretization of $x$ is less restrained, it needs to support irreversible inflow boundary conditions where the in-flowing distribution can be specified without affecting the out-flowing distribution.

\subsection{Phase space discretization}

The natural discretization method that matches the above restrictions is an upwind first order finite-difference algorithm, which is proven to be stable \cite{Frensley:1990by, Jiang:2014cb}.
However, fully coherent quantum transport implies dissipationless unitary time evolution , while finite-differences are well known to suffer from diffusive errors.
Due to the latter, unwanted smoothing or coarse-graining of the phase space variables is observed and, hence, unintentional dissipation inevitably sneaks in.
To alleviate this numerical problem, we implemented the second order MacCormack's predictor-corrector method \cite{MacCormack:1969vz}.
Unfortunately, this method introduces dispersive errors, which are found to cause numerical  instabilities when applied to the quantum generation terms.
As claimed in literature, stability could be regained by implementing a limiter \cite{Selle:2007uc}, but we found that the introduction of the latter leads to uncontrollable nonlinear errors.
Literature provides us with appropriate solutions to this problem, such as the weighted essentially non-oscillatory (WENO) method that remedies the diffusive problems related to finite differences while avoiding the dispersive errors related to oscillatory behaviour \cite{Jiang:1996eq}.
This has been successfully applied to the Wigner equation \cite{Dorda:2015gu}, albeit at the cost of a very fine grid spacing, which, as we mentioned before, is needed when the highly oscillatory Wigner kernel is to be tackled directly.
Therefore, although the finite-difference method is simple to implement at first sight, we found that good accuracy can be obtained at lower computational burden by using pseudo-spectral methods.
Pseudo-spectral methods have been successfully used to accurately treat the Wigner function in Refs.~\cite{shao:2011xx,xiong:2016xx}. However, in this section, we do not present such a highly optimized implementation. Rather, we show that the form of the Wigner function that we obtained using the spectral decomposition of the force-field allows for the use of simple deterministic numerical methods with appropriate properties as becomes clear in section \ref{implementation}.

Spectrally accurate collocation methods suppress both diffusive and dispersive errors, while still allowing to calculate the quantum generation term directly.
In particular, we applied a Chebyshev polynomial expansion in the $x$ coordinate and performed a Fourier series expansion with respect to the $p$ coordinate.
The Chebyshev polynomial and Fourier series are advantageous because both the expansion coefficients and the derivatives can be extracted from Fast Fourier Transforms (FFTs), which improves  scalability by reducing the computational complexity from $\mathcal{O}(N^2)$ to $\mathcal{O}(N\log N)$. \cite{Boyd:2001wu}
Moreover, the Chebyshev series expansion can easily accommodate inflow boundary conditions when evaluated on a Gauss-Lobatto-Chebyshev extremal grid $x_n = \cos(n\pi/N)$ for $n=0,1,...,N$.
The Wigner generation terms are evaluated directly on the phase space grid.
Because only the classical term uses the Fourier expansion, we can avoid the unphysical periodicity of momentum space by selecting a high enough maximum momentum to fully represent the Wigner function. The quantum mechanical generation terms are limited in momentum by design.

The default Chebyshev grid spacing becomes very dense at the edges of the domain as the minimal spacing scales with the inverse of the square of the number of points, $\Delta x_\mathrm{min} \sim N_x^{-2}$.
This severely limits the allowed time step, which is of the order of $\Delta t_\mathrm{max} = m\,\Delta x_\mathrm{min} /p_\mathrm{max}$.
In order to prevent this detrimental effect on the computational efficiency, we have adopted the coordinate transformation introduced by Kosloff and Tal-Ezer \cite{Kosloff:1993dr}, resulting in a more uniform spread of grid points and a linear scaling of the time step.

A well known physical feature of the ballistic Boltzmann equation (also referred to as the Vlasov equation) is the filamentation of the distribution function over time, where a particular region of phase space is stretched out over time, while ever refining its structure.
As to the Wigner function this filamentation is even more pronounced due to the quantum mechanical time evolution which readily introduces oscillations in those regions of phase space where the Wigner function turns negative.
Eventually the phase space structures become too fine for the numerical method to work out.
In low order methods, such as finite difference techniques, this leads to an unphysical appearance of dissipation because of coarse graining.
However, in spectral methods this would cause a blow-up of high frequency components due to spectral blocking, which makes the method numerically unstable.
A small redeeming quality of the spectral method is its ability to indicate incorrect results, in quite a dramatic way.
To solve this instability, we adopt a technique developed for the Vlasov equation in plasma physics referred to as filamentation filtration \cite{Klimas:1994cf}.
Avoiding a detailed discussion of this method, we would like to stress that it preserves the accuracy of the zeroth and first moments of the Wigner function, i.e. the particle density and current density.
Alternatively, we have also implemented the Hou-Li filtering method \cite{Hou:2007ec}, which is a more general anti-aliasing filter that seems to provide high accuracy in practice, even though this method does not strictly conserve accuracy of any of the moments of the Wigner function.

\subsection{Benchmarks and errors}

\begin{figure}[!b]
  \centering
  \includegraphics[width=0.45\textwidth]{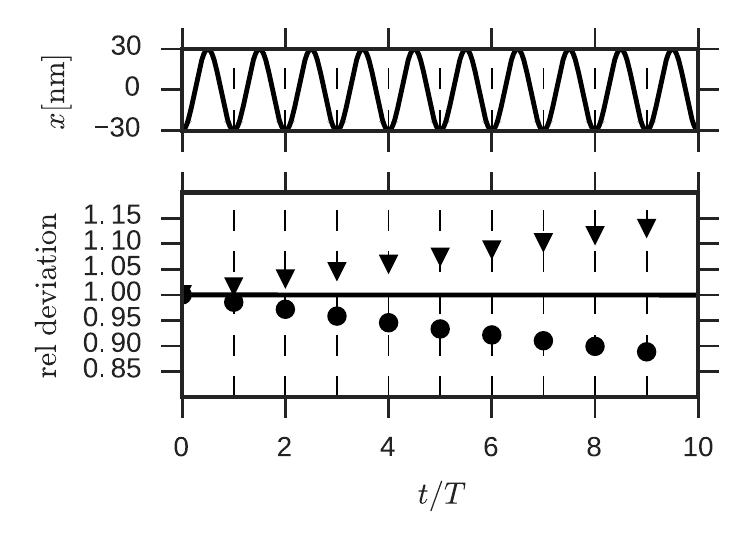}
  \caption{Time dependence of the harmonic oscillator with period $T$. Top: the mean position of the wavepacket, showing periodic oscillation between $-20$ nm and $-80$ nm. Bottom: the total density relative to the starting density (solid line), the maximum density of the wave packet at every period relative to the density at the start (circles), and the standard deviation at every period relative to the standard deviation at the start (triangles).}
  \label{f:harmonic}
\end{figure}

In this section, we verify the validity and estimate the accuracy of our approach using two test cases. Firstly, we use a wavepacket in an harmonic oscillator, with a well known analytical solution, to study the stability and error through time. Secondly, we use a double barrier potential to study the convergence of the steady state Wigner function with respect to the choice in cut-off of the wave vectors $k$ corresponding to the spectral components of the force field $F_k(x)$.

The harmonic oscillator is characterized by a quadratic potential $V(x) = \frac{1}{2}kx^2$ with force field $F(x)=-kx$, with spring constant $k=1\,\mathrm{meV}/\mathrm{nm}^2$, mass $m=0.05\,\mathrm{m_e}$. The period of this oscillator is then $T=105.94\,\mathrm{fs}$. The phase space grid has $399$ points in the position coordinate, spanning from $-100$ nm to $100$ nm, while the momentum coordinate has $301$ points, spanning from $-2/$nm to $2/$nm. The number of wave vectors $k$ is also $301$. The initial distribution is a minimally uncertain Gaussian wave packet with zero momentum located at $20\,\mathrm{nm}$ with a standard deviation of $4\,\mathrm{nm}$.

As seen in the top of Fig.~\ref{f:harmonic}, our method correctly simulates the oscillation in the harmonic potential, with the mean position of the wavepacket correctly oscillating with period $T$. In the bottom of Fig.~\ref{f:harmonic}, we observer no noticeable error in the total phase space density when evolving through time, however the wavepacket spreads out due to numerical diffusion.

\begin{figure}[!ht]
  \centering
  \includegraphics[width=0.45\textwidth]{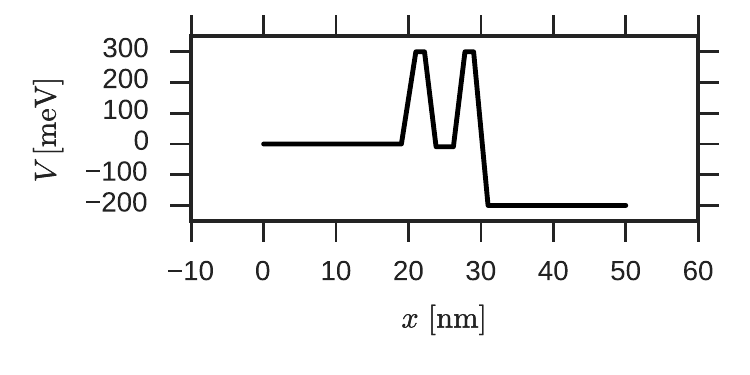}
  \caption{Double barrier potential simulated for increasing number of wave vectors $k$.}
  \label{f:doublebarrierpot}
\end{figure}

Next, we simulate a common test case: the double barrier potential shown in Fig.~\ref{f:doublebarrierpot}.

It is defined by a piecewise constant force field,
\[
  F(x) = \begin{cases}
  -0.2\,\mathrm{eV} & \mathrm{for}\ |x-20\,\mathrm{nm}|<0.75\,\mathrm{nm}\\
  \hphantom{-}0.2\,\mathrm{eV} & \mathrm{for}\ |x-23\,\mathrm{nm}|<0.75\,\mathrm{nm}\\
  -0.2\,\mathrm{eV} & \mathrm{for}\ |x-27\,\mathrm{nm}|<0.75\,\mathrm{nm}\\
  \hphantom{-}10/3\,\mathrm{eV} & \mathrm{for}\ |x-30\,\mathrm{nm}|<0.75\,\mathrm{nm}
\end{cases}\,.
\]
Note that this potential is not differentiable at the points $x=\{20, 23, 27, 30\}\,\mathrm{nm}$, therefore it does not conform to our previous definition of a smooth, ``realistic'' potential. However, we have selected this potential both for ease of reproduction and to test the behavior of our method with a worst case potential.

In these simulations, the effective mass has been set to the conduction band of InAs ($0.023\,m_0$). The simulation domain is $50\,\mathrm{nm}$ wide, discretized over $127$ points, with the momentum space spanning from $-2.5/\mathrm{nm}$ to $2.5/\mathrm{nm}$, discretized over $301$ points.
For the initial condition, as well as the injecting boundary conditions, we take the electrons to be in a Fermi-Dirac distribution with a chemical potential $\mu=100\,\mathrm{meV}$.
The time evolution is performed with steps of $0.02\,\mathrm{fs}$ up to $200\,\mathrm{fs}$, at which point the Wigner function has reached a steady state.

\begin{figure}[!ht]
  \centering
  \includegraphics[width=0.45\textwidth]{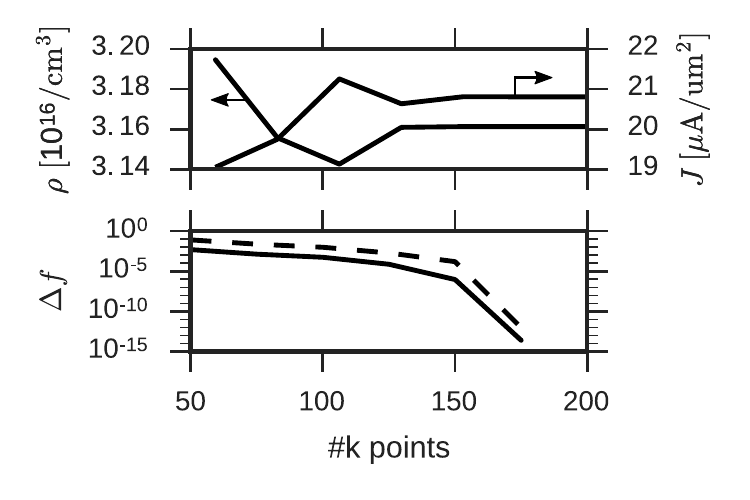}
  \caption{Simulation results of double barrier for increasing number of wave vectors $k$. Top: Particle density in the middle of the well $\rho$ and current density through the device $J$. Bottom: mean (solid) and max (dashed) error in the Wigner function relative to the result obtained using $225$ $k$-points. The result for $200$ and $225$ points is identical.}
  \label{f:doublebarrier}
\end{figure}

Fig.~\ref{f:doublebarrier} shows the resulting data for simulations using $50$ to $200$ wave vectors, in increments of $25$.
Increasing the number of points from $200$ to $225$ results in an identical Wigner function, we consider the solution to be converged at this point.
In the bottom part of Fig.~\ref{f:doublebarrier}, we compare the average and maximum errors of the Wigner function with the converged solution at $225$ points, showing rapid convergence with increasing number of wave vectors.
In the top part of Fig.~\ref{f:doublebarrier}, we show the observable density of electrons in the middle of the well at $25\,\mathrm{nm}$ and the current density through the device for a different number of wave vectors. Even at only $50$ points, the observables are already well estimated. From $150$ wave vectors on, we can no longer distinguish the resulting density and current density.

Different results are obtained when less than $150$ $k$-points are used because the limited number of spectral force components $F_k(x)$ provide too poor a representation of the applied force field. As the number of $k$-points is increased, the force field is better represented, except at the non-differential points of the potential energy where we expect Gibbs oscillations due to the numerical discretization.
However, we note that this does not affect our solution, or its convergence as long as we ensure the discretization in $k$ is sufficiently fine with a large $k$ cut-off.
To explain this, we first invoke a heuristic argument based on the underlying physics: increasing the number of $\vk$ points does not reduce the amplitude of the Gibbs oscillation fringes, but it does make them thinner. These non-physical fringes on the potential present a barrier for transport, but quantum mechanically these barriers are semi-transparent due to tunneling. As the number of $\vk$ points is increased to an arbitrary value, these fringes become more transparent until their transmission coefficient is arbitrarily close to one.
Secondly, from an analysis of (\ref{e:wl-spectral-1d}), we conclude that the effect from an omitted spectral component with large wave vector $k_\mathrm{large}$ would result in contributions to the Wigner function at momenta $p$ emerging from momenta $p\pm\hbar k_\mathrm{large}/2$, which will be negligible for large enough $k$, in particular considering the damping factor $1/\hbar k$.
Therefore, although our method is meant to be applied for sufficiently smooth potentials, it seems that even non-differentiable potential profiles which give rise to Gibbs oscillations can still produce sensible results.

In conclusion, as for any pseudo-spectral method, the accuracy of the simple Chebyshev-Fourier implementation of the spectral force-field Wigner equation we described, is limited by the smoothness of the Wigner function and the force field. Counter-intuitively, the quantum mechanical nature of the Wigner function is helpful in this context, as it  provides a smoother Wigner function than its classical counterpart.

\section{Resonant tunneling diode}
\label{rtd}

\begin{figure}[!ht]
  \centering
  \includegraphics[width=0.45\textwidth]{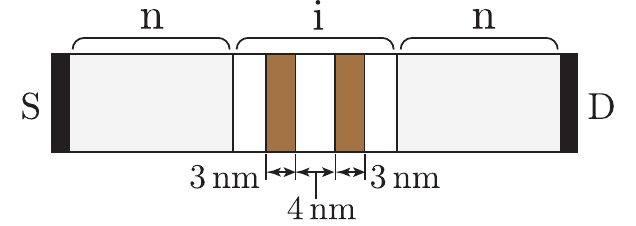}
  \caption{A schematic representation of the RTD under study. The RTD consists of two n-doped contact region with an intrinsic region in between. In the intrinsic region, two thin barriers of $3\,\mathrm{nm}$ wide with a spacing of $4\,\mathrm{nm}$  are applied by means of a band offset of $300\,\mathrm{meV}$. This structure causes the formation of resonant levels between the barriers.}
  \label{f:rtd_structure}
\end{figure}

As an illustration of a real application, we calculate the current-voltage (I-V) characteristics of a resonant tunneling diode (RTD)  \cite{Chang:1974di}.
An RTD typically exploits resonant tunneling through a double barrier as the basic mechanism leading to negative differential resistance.
In figure~\ref{f:rtd_structure}, we have schematically depicted the simulated structure and given its relevant dimensions.

For the sake of simplicity, the material parameters of InAs are used throughout the entire structure for which we have assumed translational invariance along the transverse directions.
The n-type regions have a uniform ionized donor doping concentration of $5\times10^{17}\mathrm{cm}^{-3}$.
The shaded regions in the center form the double barrier and their band-offset is assumed to be $300\,\mathrm{meV}$.
All further material and simulation parameters are listed in table~\ref{t:params}.
\begin{table}[!ht]
  \caption{Simulation and material (InAs) parameters}
  \label{t:params}
  \setlength{\tabcolsep}{8pt}
  \begin{tabular}{r|c|l}
    \hline
    electron mass & $m^*_\mathrm{e}$ & $0.023\, m_0$ \\
    heavy hole mass & $m^*_\mathrm{hh}$ & $0.41\, m_0$ \\
    light hole mass & $m^*_\mathrm{lh}$ & $0.026\, m_0$ \\
    bandgap & $E_\mathrm{gap}$ & $0.354$ \\
    dielectric constant & $\epsilon$ & $15.15\, \epsilon_0$ \\
    time step & $\Delta t $ & $0.02\, \mathrm{fs}$ \\
    \# position points & $N_x$ & $257$ \\
    \# momentum points & $N_p$ & $201$ \\
    maximum momentum & $p_\mathrm{max}/\hbar$ & $1.2\,\mathrm{nm}^{-1}$\\
    \hline
  \end{tabular}
\end{table}

\begin{figure}[!ht]
  \centering
  \includegraphics{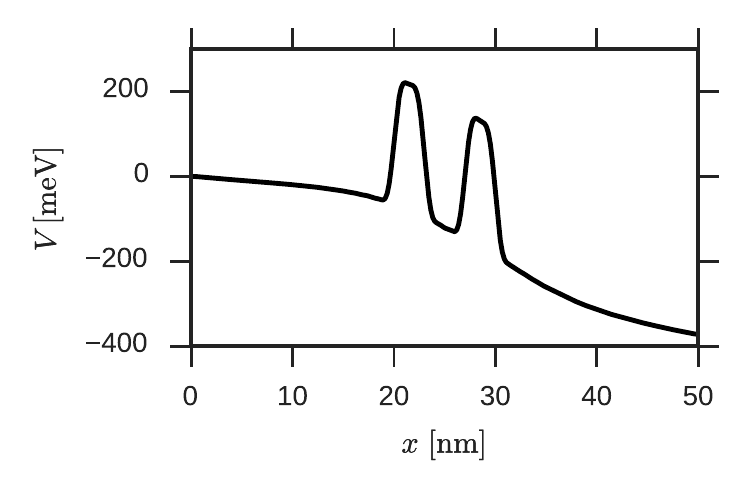}
  \caption{The self-consistently calculated potential near the resonance peak at $V_\mathrm{b}=0.4\,\mathrm{V}$.}
  \label{f:wigner_steady_on_v}
\end{figure}

The simulations cover the time evolution from equilibrium at $t=0$ with zero bias voltage towards a non-equilibrium state at a time $t > 0$ with the bias voltage $V_\mathrm{b}$ switched on.
The profile of the force field representing both electron-electron interaction in the Hartree approximation and the external bias voltage is self-consistently updated by iterating on Poisson's equation at every time step.
For reference, the self-consistent potential profile in steady state at a $0.4$ eV bias is shown in Fig.~\ref{f:wigner_steady_on_v}.

\subsection{Ballistic transport}

Because of the low switching speed, as compared to electron velocities, at which electronic devices usually operate, we are mostly interested in the steady state regime.
For the purpose of the steady state, we abruptly apply the bias voltage $V_\mathrm{b}$ at $t=0$.
Corresponding formally to the limit $t\to\infty$, the steady state is attained once the RTD current doesn't appreciably vary with time any longer.
For the RTD under consideration, the typical settling time is on the order of $1\,\mathrm{ps}$, or about $50.000$ numerical time steps.

\begin{figure}[!ht]
  \centering
  \includegraphics{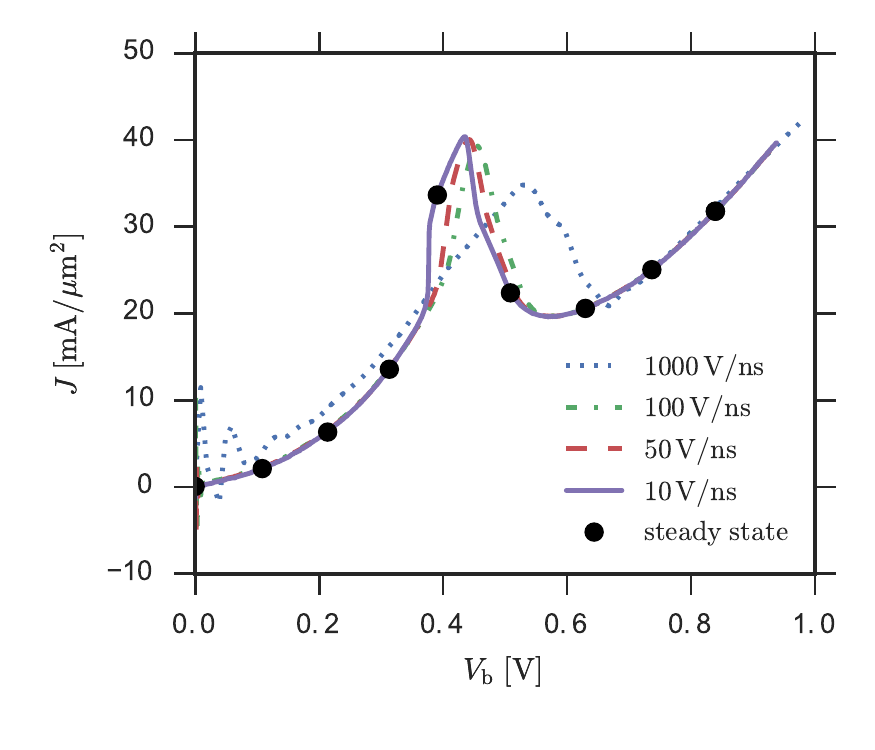}
  \caption{I-V characteristic of the RTD. The dots represent the conventional steady-state I-V curve, while the dashed and full line are the results of a time-dependent linear bias sweep with different rates as given in the legend.}
  \label{f:iv}
\end{figure}

In Fig.~\ref{f:iv} we show the ballistic I-V characteristic in dots, where every point represents a different steady state simulation.
As expected, the current peak is reached when the resonant level in the well aligns with the injected electron distribution, while at lower and higher bias, the mismatch makes the current decrease.
This quantum effect causes a region of negative differential resistance (NDR) to occur beyond the resonance peak.
At even higher currents, ballistic emission of electrons surmounting the barriers again increases the current.

\begin{figure}[!ht]
  \centering
  \includegraphics{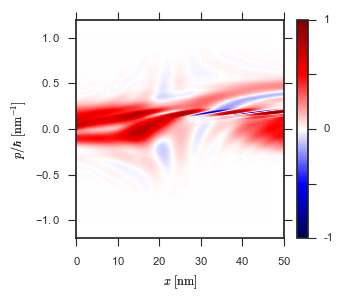}
  \caption{The Wigner function near the resonance peak at $V_\mathrm{b}=0.4\,\mathrm{V}$ in arbitrary units. (color online)}
  \label{f:wigner_steady_on}
\end{figure}

For the sake of reference, we have shown the Wigner function near resonance in Fig.~\ref{f:wigner_steady_on}, corresponding to the potential profile in Fig.~\ref{f:wigner_steady_on_v}.
One clearly sees ``negative regions'' where the Wigner function takes negative values, reflecting non-classical behaviour \cite{Kenfack:2004cv}.
These negative regions sustain a current, carried by a jet of positive quasi-probability density electrons, that tunnel through the resonant level and is accelerated on the right hand side of the barriers.

In Fig.~\ref{f:iv}, we have also plotted the results of several sweeps of the applied bias over time.
These sweeps are set-up as a linear increase in applied bias from $0\,\mathrm{V}$ to $1\,\mathrm{V}$ at different rates, ranging from $10\,\mathrm{V/ns}$ to $1000\,\mathrm{V/ns}$.
Because the system has not yet attained a steady state under these conditions, the current density is generally not uniform throughout the RTD structure.
The current density shown in Fig.~\ref{f:iv}, is thus the current density flowing into the device at the left hand side.

At higher sweep rates, in particular at $1000\,\mathrm{V/ns}$, the sweeps become too fast for the electrons to keep up.
We can still discern a typical RTD characteristic, but it does not match with the steady state points.
Moreover, we observe a significant lag of the resonance peak as compared to the steady state case.
On the other hand, at $10\,\mathrm{V/ns}$, the slowest rate we have simulated, the I-V curve entirely coincides with the steady state points.
The ability to fully reproduce the steady-state current indicates that this sweep rate can be considered adiabatic.
Even more evidence for the latter is provided by the current density remaining constant throughout the structure at every time point (not shown here for the sake of brevity).
By performing an adiabatic sweep, we obtain a very detailed I-V characteristic in a single time-dependent simulation as the current value at any time point can be used.
In our example, we are dealing with over a million time-steps, and can thus obtain the I-V characteristics with a resolution below $1\,\mathrm{\mu V}$.
This greatly extends the utility of time-dependent calculations to probe the steady-state.

\begin{figure}[!ht]
  \centering
  \includegraphics{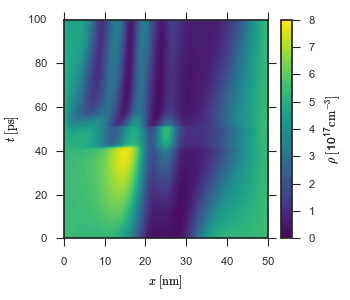}
  \caption{Density profile at the different time points of the $10\,\mathrm{V/ns}$ sweep. (color online)}
  \label{f:density_time}
\end{figure}

The increased resolution of the bias sweep further reveals the fine structure of the resonance peak, which wasn't to be expected at first.
We observe a sharp increase in current just before resonance sets in at about $380\,\mathrm{mV}$, as well as a sharp decrease immediately following the resonance peak near $440\,\mathrm{mV}$.
To explain these phenomena, we have plotted the electron density profile along the RTD structure versus position and time in Fig.~\ref{f:density_time}.
We observe that that the sharp transitions in the I-V characteristic correspond to electrons occupying the resonant level.
The resonant state in the well always gets partially populated by tunneling through a single barrier.
However, once resonant current can flow, the density in the well drastically increases.
Getting occupied, the resonant state level pushes itself to higher energy, in line with the doped left contact, which is able to sustain the occupation of the resonant level.
The corresponding amplification effect survives until the bias exceeds the charge density capacity of the well.
Then, the state drops in energy, gets abruptly depleted and does not carry any more current, all of which explains the sharp transitions observed.

\subsection{Dissipative transport}

\begin{figure}[!ht]
  \centering
  \includegraphics{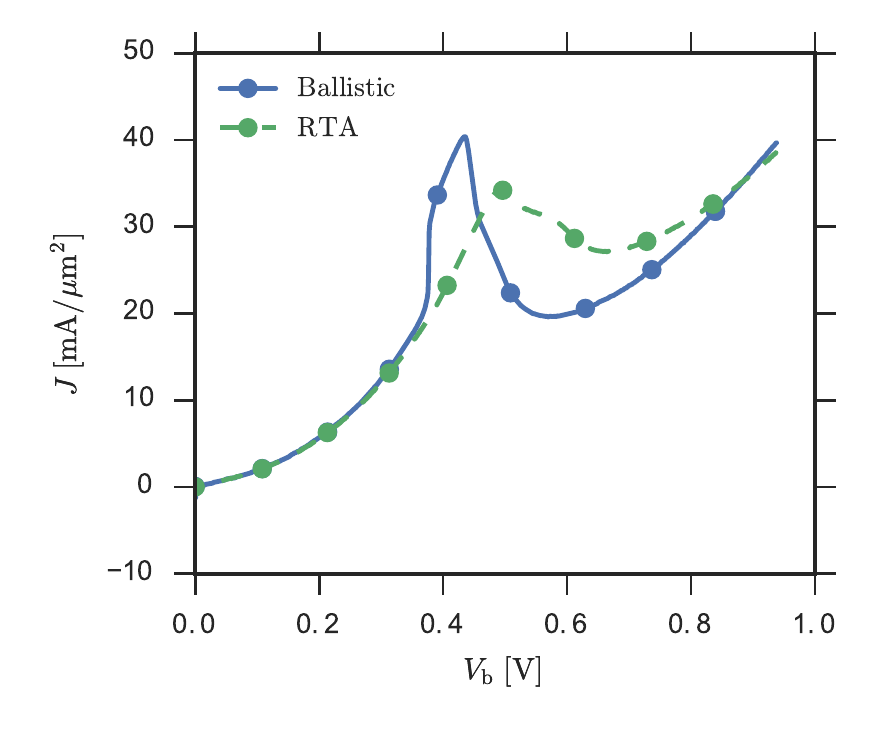}
  \caption{I-V characteristics of the RTD with and without RTA ($\tau=100\,\mathrm{fs}$). As in Fig.~\ref{f:iv}, dots are steady-state points, whereas the full and dashed lines are linear bias sweeps at a rate of $10\,\mathrm{V/ns}$.}
  \label{f:iv_relax}
\end{figure}

In realistic devices, electrons cannot propagate exclusively in ballistic mode.
Instead they interact with their environment, thus being subjected to energy dissipation.
In this example, we have included energy dissipation within the relaxation time approximation (RTA).
We arbitrarily chose a short relaxation time $\tau=100\,\mathrm{fs}$, which will serve to show the effects of scattering, even with a very low effective mass.
In Fig~\ref{f:iv_relax}, we compare the results of the RTA to the previously obtained ballistic currents.

\begin{figure}[!ht]
  \centering
  \includegraphics{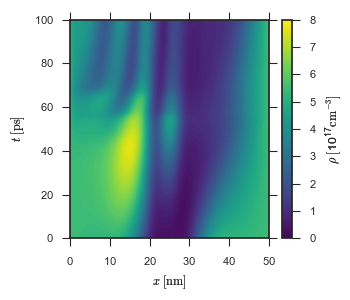}
  \caption{Density profile at the different time points of the $10\,\mathrm{V/ns}$ sweep with dissipation using RTA. (color online)}
  \label{f:density_time_rta}
\end{figure}

The addition of energy relaxation changes the I-V characteristics in many ways, and we have listed the most prominent effects below.
Before the onset of resonance, the current is lower than its ballistic counterpart due to resistive backscattering of electrons, i.e. the addition of a resistive component.
The resonance peak is lowered, as electrons may be scattered out of dominant resonant level.
The peak itself is shifted towards a higher bias voltage, partly due a voltage drop over the resistive regions emerging in front of the barriers, but more importantly because inelastic scattering may enhance the occupation of the resonant level, thereby increasing its energy in line with the electrostatics, as can be observed from Fig.~\ref{f:density_time_rta}.

\section{Conclusions}
\label{conclusions}

We have derived a new form of the Wigner-Liouville equation based on the spectral components of the real space force field instead of the potential profile. With this new form it is possible to interpret the quantum mechanical evolution of a phase space distribution function as a classical evolution driven by a space averaged force field combined with the local generation of particles contributing neighboring momentum states. This interpretation is the continuum equivalent of the generation of positive and negative particles outlined in the Wigner Monte Carlo method \cite{Nedjalkov:2008ed,Sellier:2013ga}.

We have implemented the spectral force based method in a time-evolution solver using Chebyshev and Fourier pseudo-spectral methods.
As the latter invoked fast Fourier transforms, excellent spectral accuracy at reduced computational cost was obtained.
Substantial robustness of the implementation is provided by adding spectral filtering, which also preserves the relevant moments of the distribution function.

We have shown that our method can be used to calculate both the transient response and the steady-state current flowing through an RTD diode, incorporating the typical resonance peak and the negative differential resistance region.
Applying slow, adiabatic bias sweeps, we were able to get very high resolution I-V characteristics.
In addition, exploiting the high resolution resulting from suitable bias sweeps, we have attributed the observed pinning of the resonant current to an interplay between the sharp resonant level and the corresponding electrostatic features.

Finally, the incorporation of inelastic scattering through a relaxation-time approximation reduces the peak-to-valley ratio of the RTD current, while also the time required to attain the steady state was found to be decreased, due to the quantum mechanical overshoot being damped.

\bibliography{spectral-force-wigner.bib}

\end{document}